\documentclass{emulateapj}
\usepackage{times}

\newcommand{\HST}{{\sl HST}}

\newcommand{\VI}{$V\!-\!I$}

\newcommand{\picplace}[1]{\vbox{\hrule\@height 0.4pt\@width\hsize
\hbox to\hsize{\vrule\@width 0.4pt\@height#1\hfil
\vrule\@width 0.4pt\@height#1}\hrule\@height 0.4pt\@width\hsize}}

\slugcomment{Accepted by ApJ Letters}

\shorttitle{Dynamical evolution of second-generation globular clusters in 
  NGC~1316}
\shortauthors{Goudfrooij et al.}

\begin{document}


\title{Deep Luminosity Functions of Old and Intermediate-Age Globular
  Clusters in NGC~1316: Evidence for Dynamical Evolution of
  Second-Generation Globular Clusters\altaffilmark{1}} 


\author{Paul Goudfrooij, Diane Gilmore, and Bradley C. Whitmore}
\affil{Space Telescope Science Institute, 3700 San Martin Drive,
  Baltimore, MD 21218} 
\email{goudfroo@stsci.edu}

\and 

\author{Fran\c{c}ois Schweizer} 
\affil{Carnegie Observatories, 813 Santa Barbara Street, Pasadena, CA 91101}


\altaffiltext{1}{Based on observations with the NASA/ESA {\it Hubble
    Space Telescope}, obtained at the Space Telescope Science
    Institute, which is operated by AURA, Inc., under NASA contract
    NAS5--26555.} 


\begin{abstract}
The Advanced Camera for Surveys on board the {\it Hubble Space
    Telescope\/} has been used to obtain deep high-resolution images
    of the giant early-type galaxy NGC~1316 which is an obvious
    merger remnant. These observations 
    supersede previous, shallower observations 
    which revealed the presence of a population of metal-rich globular clusters
    of intermediate age ($\sim$\,3 Gyr).  
    We detect a total of 1496 cluster candidates, almost 4 times as many as
    from the previous WFPC2 images.  
    We confirm the bimodality of the color distribution of clusters,
    even in \VI, with peak colors 0.93 and 1.06. 
    The large number of detected clusters allows us to evaluate the globular
    cluster luminosity functions as a function of galactocentric radius. 
    We find that the luminosity function of the inner 50\% of the
    intermediate-age, metal-rich (`red') population of clusters differs
    markedly from that of the outer 50\%. In particular, the luminosity
    function of the inner 50\% of the red clusters shows a clear 
    flattening consistent with a turnover that is about 1.0 mag fainter than the
    turnover of the blue clusters. This constitutes the first direct evidence that
    metal-rich cluster populations formed during major mergers of gas-rich
    galaxies can evolve dynamically (through disruption processes) into the red,
    metal-rich cluster populations that are ubiquitous in `normal' giant
    ellipticals.  

\end{abstract}




\section{Introduction}              \label{s:intro}
Recent deep imaging studies of `normal' giant elliptical galaxies with the
{\it Hubble Space Telescope (HST)\/} and large, ground-based telescopes
have shown that these galaxies usually contain rich globular cluster
(GC) systems with bimodal color distributions (e.g.,
\citealp*{kunwhi01,lars+01}). Typically, roughly half of the GCs
are blue, and half red. Follow-up spectroscopy with 8-m class telescopes
revealed that both `blue' and `red' GC subpopulations are typically old
($\ga$\,8 Gyr, \citealp*{forb+01,cohe+03}) implying that the
bimodality is mainly due to differences in metallicity. The colors of the
`blue' GCs are usually similar to those of metal-poor halo GCs in the
Milky Way and M31, while the mean colors of the `red' GCs are
similar to those of the diffuse light of their host galaxies (e.g.,
\citealp*{geis+96,forb+97}). 
Hence, the nature of the `red', metal-rich GCs is likely to hold important
clues to the star formation history of their host galaxies. 

One environment {\it known\/} to produce metal-rich GCs and bimodal color
distributions is that of vigorous starbursts induced by mergers of gas-rich
galaxies. Massive young GCs have been commonly found in
mergers and young merger remnants using {\it HST\/} observations (e.g.,
\citealp*{holt+92,schw02}, and references therein). 
Follow-up spectroscopy has confirmed the ages 
(and in one case even the high masses,  \citealp*{mara+04}) 
of these young clusters predicted from their
colors and luminosities (e.g., \citealp{zepf+95,schsei98}). Their metallicities
tend to be near solar, as expected for clusters formed out of enriched gas
in spiral disks.  
A natural interpretation of these data is that the metal-rich GCs in 
`normal' giant ellipticals formed in gas-rich mergers at $z \ga 1$, and that the
formation process of giant ellipticals with significant populations of
metal-rich GCs was similar to that in galaxy mergers observed today
\citetext{e.g., \citealp*{schw87,ashzep92}}. 

However, one important, hitherto unsurmounted hurdle for this `formation by
merger' scenario has been the marked difference in the luminosity functions
(hereafter LFs) of old vs.\ young GC systems (e.g., \citealp{vdb95a}).  
The LFs of old GC systems of `normal' galaxies are well fit by 
Gaussians peaking at $M_V^0 \simeq -7.2$ mag with a dispersion of
$\sigma \simeq 1.3$ mag \citetext{e.g., \citealp{whit97}}, while young GC
systems in mergers and young remnants have power-law LFs with indices of
$\alpha \simeq -2$ \citetext{e.g., \citealp{whit+99}}. 
Recent calculations of 
dynamical evolution of GCs
(including two-body relaxation, tidal shocking, and stellar mass loss) show
that the least massive GCs disrupt first as galaxies age, which can  
gradually transform the initial power-law LFs into LFs with Gaussian-like
peaks or turnovers \citetext{e.g., \citealp{falree77,baum98,falzha01}; but see 
  \citealp{vesp01}}.  
Observational evidence of this effect has been reported for 
a GC system near the center of M\,82, featuring 
a very short GC disruption timescale 
(de Grijs, Bastian, \& Lamers \citeyear{degr+03}).  

Intermediate-age merger remnants with ages of 1\,--\,5 Gyr
are ideal probes for studying the long-term dynamical effects on GC systems
formed during a major merger. Such galaxies are still 
identifiable as merger remnants through their morphological fine structure 
\citetext{e.g., \citealp{schsei92}}, 
yet are old enough to ensure that substantial dynamical evolution of
the globular clusters has already occurred. 

Recent \HST\ studies of candidate intermediate-age merger remnants have
revealed that their `red' GC subpopulations show LFs consistent with power 
laws (as expected if formed during a recent merger event;
\citealp*{goud+01b,whit+02}). However, the completeness limits of those
studies were not faint enough to allow a detection of a turnover in the
LFs. This {\it Letter\/} reports on new observations of the intermediate-age
merger remnant NGC~1316 using the Advanced Camera for Surveys {\it (ACS)\/},
installed on {\it HST\/} in March 2002, whose unprecedented sensitivity
allows us, for the first time, to detect these turnovers. 

\section{NGC~1316}              \label{s:n1316}

NGC~1316 is a prime example of an inter\-medi\-ate-age merger
remnant. Extensive optical observations showed that NGC~1316 is a typical
Morgan D-type galaxy with a surface brightness profile following an $r^{1/4}$
law \citep{schw80,schw81}. Its outer envelope includes several non-concentric
arcs, tails and loops that are most likely remnants of tidal perturbations,
while the  inner part of the spheroid is characterized by a surprisingly high
central surface brightness and small effective radius for the 
galaxy's luminosity. All of these features are consistent with NGC~1316
being the product of a dissipative merger with incomplete dynamical
relaxation. Recently, Goudfrooij et al.\ \citeyearpar{goud+01a,goud+01b}
discovered a significant population of $\sim$\,3 Gyr old GCs of near-solar
metallicity through a comparison of {\it BVIJHK\/} colors as well as H$\alpha$
and Ca\,{\sc ii}-triplet line strengths with population synthesis model
predictions.   

In the following we adopt a distance of 22.9 Mpc 
for NGC~1316 \citetext{see \citealp*{goud+01b}, hereafter Paper I}.

\section{Observations and Data Analysis}

NGC~1316 was observed with {\it HST\/} on March 4 and 7, 2003, using the
wide-field channel of {\it ACS\/} and the F435W, F555W, and F814W filters, with
total exposure times of 1860 s, 6980 s, and 4860 s, respectively. 
In this {\it Letter\/} we concentrate on the F555W and F814W images, which
reach significantly deeper than the F435W image. 
The final F555W and F814W images were constructed from long (18\,--\,21
min) exposures at 4 or 6 dither positions, supplemented by a few
short exposures to avoid saturation of the central regions. 
The individual images in each band were combined using the task {\sc
  multidrizzle} within {\sc iraf/stsdas v3.2}. 
Saturated pixels in the long exposures were replaced by those in the short
exposures while running {\sc multidrizzle} by setting the appropriate data
quality flag in the affected pixels of the long exposures.  
The high sensitivity of the {\it ACS\/} images
allowed us to reach about 1.9 mag deeper than the {\it WFPC2\/} images
used in Paper~I did. 

WFPC2 images were taken in parallel to the ACS observations, $\sim$\,5\farcm5 away
from the ACS pointings, in order to provide estimates for the number density of
compact background galaxies and foreground stars as a function of brightness
and color (see Sect.~\ref{s:LFs}). The
WFPC2 exposures consisted of a total of 5100 s in F555W and 3000 s in
F814W. Each set of WFPC2 images was also combined using {\sc multidrizzle}.   

The cluster selection procedure followed the one described in 
Paper I, to which we refer the reader interested in details. 
Briefly, we first built smooth galaxy model images 
from the combination of an isophotal model composed of pure ellipses and a
median-filtered version of the image. The objects were then selected by
applying the {\sc daofind} task from the {\sc daophot\/} package
\citep{stet87} to an image prepared by dividing the F555W image by the square
root of the corresponding model image (thus having uniform shot noise
characteristics over the whole image). The detection threshold was set at
3\,$\sigma$ above the background. Detections located  
within or on the edge of the dust features in NGC~1316 were excluded from
the sample. 
As the resolution of the drizzled ACS images is high enough to resolve
typical GCs at the distance of NGC~1316, a 
further subselection was made by restricting a `concentration
index' (the difference in magnitudes measured using radii of 2 and 5 pixels)
to stay between 0.35 and 0.75 mag. This effectively filtered out extended
background galaxies as well as bright foreground stars and any remaining hot
pixels.  
Aperture corrections were $-0.49 \,(\pm0.01)$ mag in $V$ and $-0.56 \,(\pm0.01)$
mag in $I$ for an 
aperture radius of 3 WFC pixels (i.e., 0\farcs15). These values were determined
through measurements of several isolated, well-exposed GC candidates located
throughout the field.  
The transformation of the instrumental magnitudes to 
Johnson/Cousins $V$ and $I$ was performed using the
calibrations derived by the ACS Instrument Definition Team (Sirianni et
al., in preparation).  

\begin{figure}
\epsscale{0.85}
\plotone{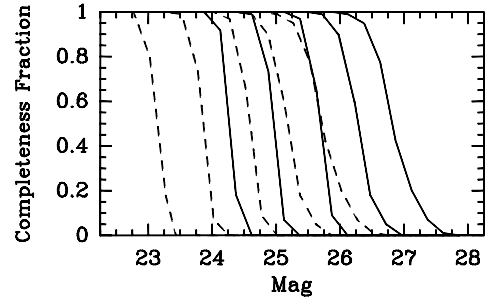}
\caption{Completeness functions used for the ACS/WFC photometry of
NGC~1316 GC candidates. Lines (solid for $V$ band, dashed for $I$ band) 
are shown for five background levels: From left to right: 
1600, 800, 320, 200, and 140 e$^{-}$ 
pixel$^{-1}$ for an exposure time of 1100 s.\label{f:completeness}}  
\end{figure}

Completeness corrections were performed for five background levels and several
magnitude intervals by adding artificial objects (in batches of 300) with a
radial intensity profile derived from a fit to real GC candidates in each frame
(using {\sc daophot} routines). The
object colors were set equal to the median color of GC candidates. Figure
\ref{f:completeness} shows the resulting completeness functions.   

\section{Two Subpopulations and Their Luminosity Functions} \label{s:LFs}

\begin{figure}
\epsscale{1.1}
\plotone{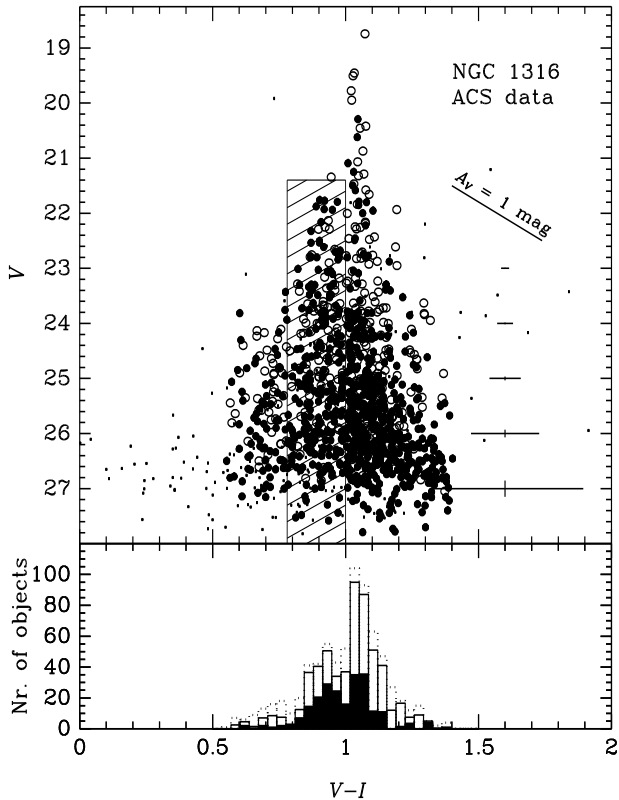}
\caption{{\it Top panel}: $V$ vs.\ \VI\ color-magnitude diagram
  for compact sources in NGC~1316. Circles represent GC candidates, while
dots represent foreground star candidates, i.e.\ objects with concentration
parameters (see text) that are consistent with that of the ACS/WFC point spread
function. The inner 50\% of the GC candidates are shown as open circles,
while the outer 50\% are shown as filled circles.  
The hatched region represents the magnitude and color
range for GCs in the Milky Way halo, placed at the distance of NGC~1316. 
Representative error bars are shown on the right-hand side of the diagram.  The
tilted line shows the reddening vector for $A_V$ = 1 mag of extinction. {\it
  Bottom panel}: \VI\ color distribution for GC candidates in NGC~1316, showing
  all GCs with $V \leq 26$ (open histogram), and GCs with 
  $V \leq 24.5$ (filled histogram). The dashed lines represent GCs 
  with $V \leq 26$ before correction for contamination by compact
  background galaxies.\label{f:cmdhist}}    
\end{figure}

The top panel of Figure~\ref{f:cmdhist} depicts the 
$V$ versus \VI\ color-magnitude diagram (hereafter CMD) for GCs in NGC~1316. 
The \VI\ colors of the GCs that are more luminous than any GC in the Milky Way
are quite uniform and lie somewhat redward of the range covered by  
the Milky-Way halo GCs. \citet{goud+01a} analyzed spectra of three of these
bright GCs, and found all three to be $\sim$\,coeval with an age of 3.0 $\pm$
0.5 Gyr and a solar metallicity (to within $\pm$\,0.15 dex).  
The fainter part of the CMD reveals a population of GCs with colors and
luminosities that are consistent with those of halo GCs in our Galaxy
\citetext{taken from the database of \citealp*{harr96}}, as well as a  
larger population with a mean color that is consistent with those of the
aforementioned brightest, metal-rich GCs throughout the sampled range of
luminosities. 
As to the potential effect of dust extinction within NGC~1316,
our selection criteria 
seem to have rendered the number of GC
candidates with colors redder than expected (given the photometry
errors) negligible regarding the discussion below. 

The \VI\ color distribution of the GC candidates is shown in the bottom panel
of Figure~\ref{f:cmdhist}, for two brightness cuts. The bimodality appears 
more pronounced than in Paper I, and is clearly present 
down to $V$ = 26 where the typical uncertainty in \VI\ is $\pm$\,0.12
mag. Prior to creating this 
diagram, the color histogram of candidate GCs was corrected for contamination by
compact background objects. This correction was derived from the
list of compact objects detected in the three WF chips of the parallel,
drizzled WFPC2 images, selected according to the same criteria as the GC
candidates in the ACS images. The color distribution of the objects in the
WFPC2 images was smoothed with a three-bin kernel to diminish small-number
noise and scaled to the area of the ACS image prior to subtraction from the ACS
color histogram. The effect of this correction 
is shown in Figure~\ref{f:cmdhist}.  

\begin{figure*}
\epsscale{1.15}
\plotone{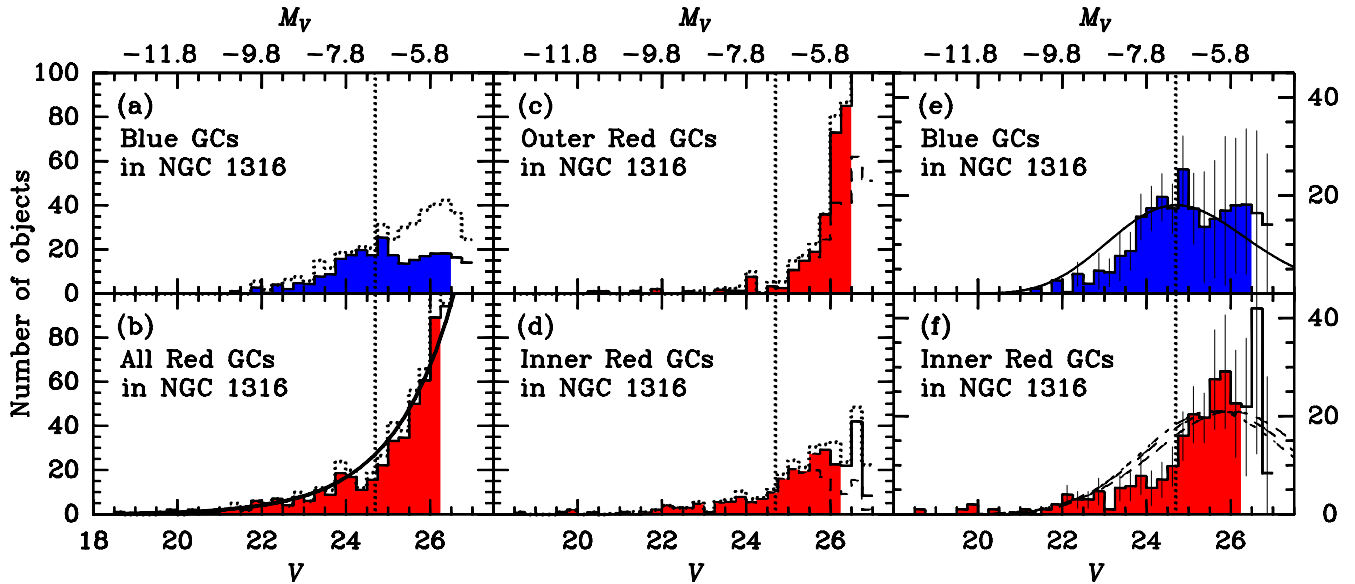}
\caption{$V$-band luminosity functions (LFs) of GC candidates in NGC~1316. 
  {\it Panel (a)}: LF of the full blue subsystem. 
  {\it Panel (b)}: LF of the full red subsystem. 
  {\it Panel (c)}: LF of the outer 50\%\ of the red subsystem. 
  {\it Panel (d)}: LF of the inner 50\%\ of the red subsystem. 
  Dashed histograms [only in panels (c) and (d)] mark uncorrected (observed)
  LFs, dotted histograms mark LFs corrected for incompleteness, and solid
  histograms mark LFs corrected for completeness and background galaxies. 
  Histograms are filled for magnitude bins brighter than the overall 50\%\
  completeness limit of the sample plotted, and open beyond 
  it. The smooth curve in panel (b) is a power-law fit to the LF. The dotted
  vertical lines represent the predicted turnover magnitude for `old' GC
  systems (i.e., $M_V = -7.2$). 
  Note the appearance of a flattening consistent with a turnover in the LF of
  the inner 50\% of the red subsystem, which is not present in the outer 50\%. 
  {\it Panel (e)}: Same LF as in panel (a), but at a different scale and with
  formal errorbars.  
  {\it Panel (f)}: Same LF as in panel (d), but at a different scale and with
  formal errorbars.  
  The smooth curves in panels
  (e) and (f) represent predicted LFs based on a combination of the Fall \&
  Zhang (2001) GC disruption models and the population synthesis models of
  \citet{mara+01}. Curves are drawn for the following populations:
  ages of 1.5 (short-dashed line), 3.0 (long-dashed line), 6.0 Gyr (dash-dotted
  line), and 12 Gyr (solid line in panel (e)). $Z/Z_{\odot}$ = 0.02 is assumed for
  the 12 Gyr curve, whereas $Z/Z_{\odot}$ = 1.0 is assumed for the three younger
  ages. The peak amplitudes of all model curves in panels (e) and (f) are 
  scaled to coincide with those of the underlying histograms. 
\label{f:LFs}}    
\end{figure*}

The luminosity functions of the blue and red GC subpopulations are shown in
Figure~\ref{f:LFs}, and were determined as follows. We first separated the GC
candidates into `blue' ($0.55 \le V\!-\!I \le 0.97$) and `red' ($1.03 \le
V\!-\!I \le 1.40$) subsamples. We avoided the overlapping region ($0.97 <
V\!-\!I < 1.03$) to minimize the mixing of the two subsamples. 
A completeness correction was then applied to each GC by dividing 
its count of 1 by the completeness fraction corresponding to its brightness and
background level. The latter fraction was calculated from the functions shown
in Fig.~\ref{f:completeness}, using bilinear interpolation in log(background)
space.  
The LFs were further corrected for contamination by compact background objects
using the smoothed and scaled LF of the objects in the parallel WFPC2 images,
as described in the previous paragraph. 
The corrected LFs 
are shown as solid lines in Fig.~\ref{f:LFs}, whereas the raw, `observed' LFs
are shown as dotted lines.  

As found in Paper I, the LFs of the (full) blue and red GC subsystems in
NGC~1316 are dramatically different from each other. The shape of 
the LF of the blue GC system is consistent with that for GCs in `normal'
galaxies and peaks near the expected turnover magnitude of $M_V$ = $-$7.2. In
contrast, the LF for the red GC system is consistent with a power law
($\phi(L)\,dL \propto L^{\alpha}\,dL$ with $\alpha = -1.75 \pm 0.07$) down to 
the 50\% completeness limit of the data. At face value, these results do not
provide significant evidence to suggest an evolution of the LF from a power law
to a Gaussian. 

However, the disruption timescales of dynamical effects 
thought to be responsible for the Gaussian shape of the GC LFs of `normal'
galaxies through preferential depletion of low-mass GCs 
\citetext{e.g., \citealp*{falzha01}, hereafter FZ01} depend on
galactocentric distance. The stronger tidal field in the central regions does not
only yield stronger tidal shocks than in the outskirts, it also imposes a more
stringent 
zero-energy surface on the GCs so that stars can be more easily
removed from the GCs through two-body relaxation. 
Hence, the appearance of a turnover in the GC LF
can be expected to first occur in the inner parts of galaxies 
\citetext{see also  \citealp*{gned97}}.  

With the 1496 GC candidates in our ACS images of NGC~1316, we are now in a
position to test this idea for a $\sim$\,3 Gyr-old merger remnant with 
adequate 
statistical significance.  To this end, we 
divide the red GC system into two equal-size parts sorted by projected
galactocentric radius. For our sample, this division ends up at 85\farcs2,
or 9.4 kpc from the galaxy center. The result is shown in 
Fig.~\ref{f:LFs}(c) and (d): The LF of the {\it outer\/} half of the red GC system
still rises 
all the way to the detection limit, but {\it the LF of the
  inner half of the red GC system shows a clear flattening consistent with a
  turnover at $\sim$\,1.0 mag fainter than that of the blue GC system}.  

We compare these results with predictions of GC disruption models by
FZ01 in Fig.~\ref{f:LFs}(e) and (f). The FZ01 
models, whose initial GC velocity distribution involves a radial anisotropy
which is similar to that of halo stars in the solar neighborhood,  
show that an initial power-law or Schechter mass function evolves in 12 Gyr to
a peaked mass distribution similar to that observed for the Milky Way GC
system. To predict the shapes of {\it luminosity\/} functions as a function of
age, we use the FZ01 model that uses a Schechter initial mass function in
combination with the stellar evolution models of \citet{mara+01}. We use
$Z/Z_{\odot}$ = 1.0 for populations of ages 1.5, 3, and 6 Gyr, while we use
$Z/Z_{\odot}$ = 0.02 for the 12 Gyr old population, similar to the median
metallicity of halo GCs in our Galaxy. 
Prior to plotting, all resulting LFs were shifted by +0.2 mag so as to
let the peak luminosity for the 12 Gyr old, metal-poor model be at $M_V = -7.2$.
As Fig.~\ref{f:LFs}(f) shows, the location of the turnover found for
the inner half of the red GC system is in remarkably good agreement with those
of the FZ01 models for intermediate ages and solar
metallicity. It thus seems reasonable to assume that the LF of this
second-generation GC system will evolve further to become consistent with the
LFs of red GCs in `normal' giant ellipticals. 

\section{Conclusion}   \label{s:conc}

We have presented the discovery of a clear flattening consistent with a turnover in
the LF of the inner half of the red, metal-rich GC subsystem of NGC~1316, a
$\sim$\,3 Gyr-old merger remnant. The turnover magnitude is consistent with
predictions of the \citet{falzha01} models at that age and metallicity. The
implication is that the dynamical evolution of metal-rich GC populations
formed in gas-rich galaxy mergers {\it can\/} change their properties so 
that they become consistent with the properties of the red, metal-rich GC
subsystems that are ubiquitous in `normal' giant ellipticals. This discovery
weakens the one remaining argument against the `formation by merger' scenario
for the metal-rich GC subsystems in `normal', giant ellipticals, namely that
the LFs of GC systems in young and intermediate-age merger remnants are
inconsistent with the Gaussian LFs of old systems. Our results support a picture
in which the formation process of giant ellipticals with significant populations
of metal-rich GCs was similar to that in gas-rich galaxy mergers observed today.

\paragraph*{Acknowledgments.}~We thank the anonymous referee for a very 
constructive review. Support for {\it HST\/} Program GO-9409 was
provided by NASA through a grant to PG from the Space Telescope
Science Institute, which is operated by the Association of Universities for
Research in Astronomy, Inc., under NASA contract NAS5--26555.

\end{document}